\documentclass[doublecol]{epl2} 
\usepackage{psfrag}
\usepackage{epsfig}

\usepackage{graphicx}
\usepackage{subfigure}
\usepackage{dcolumn}
\usepackage{bm}
\usepackage{pst-node}
\usepackage{psfrag}
\usepackage{epsfig}

\newcommand     {\beq}[1]         { \begin{equation} #1 \end{equation} }

\title{Disorder induced brittle to quasi-brittle transition in fiber bundles}
\shorttitle{Disorder induced transition in fiber bundles}

\author{Ehud Karpas and Ferenc Kun}
\shortauthor{E.\ Karpas and F.\ Kun}
\institute{                    
  \inst{}
Department of Theoretical Physics, University of Debrecen, P.\
O.\ Box:5, H-4010 Debrecen, Hungary}

\date{\today}

\abstract{
We investigate the fracture process of a bundle of fibers with random Young modulus 
and a constant breaking strength. For two component systems we show
that the strength of the mixture is always lower than the strength of the individual 
components.
For continuously distributed Young modulus the tail of the distribution
proved to play a decisive role since fibers break in the decreasing order
of their stiffness. Using power law distributed stiffness values we demonstrate
that the system exhibits a disorder induced brittle to quasi-brittle transition 
which occurs analogously
to continuous phase transitions. Based on computer simulations
we determine the critical exponents of the transition and construct the phase diagram of the system.
}
\begin{document}
\maketitle

\section{Introduction}
The fracture of heterogeneous materials is a very important scientific problem
with a broad spectrum of technological applications 
\cite{zapperi_alava_statmodfrac,pradhan_failure_2010}. During the past decade large efforts have
been devoted to obtain a deeper understanding of the role of disorder 
in fracture processes. On the one hand, the problem is very important from a 
technological point of view
in order to be able to design novel type of composite materials with high
strength at a low weight. On the other hand, due to the decisive role of disorder,
fracture phenomena address interesting challenges also for statistical physics 
\cite{zapperi_alava_statmodfrac,pradhan_failure_2010}. 

One of the most important modeling approaches to the fracture of heterogeneous
materials is the fiber bundle model (FBM) \cite{pradhan_failure_2010, 
sornette_prl_78_2140,kloster_pre_1997,naoki_prl2008,raischel_pre_2006,kun_basquin_prl2008,
hansen_crossover_prl,kovacs_mix_gls}. 
In the framework of the model the sample is discretized in terms 
of a parallel bundle of fibers. The system is typically loaded
parallel to the fiber direction by slowly increasing the external load
or the deformation. Assigning appropriate physical properties
to fibers such as Young modulus, breaking strength, and rheological behaviour, 
several important features 
of the fracture process can be reproduced. During the last two decades FBMs have
provided
a very useful insight into the fracture of heterogeneous media both 
on the macroscopic and microscopic length scales \cite{pradhan_failure_2010, 
sornette_prl_78_2140,kloster_pre_1997,naoki_prl2008,raischel_pre_2006,kun_basquin_prl2008,
hansen_crossover_prl,kovacs_mix_gls}. 
A very important common feature
of all these investigations is that the Young modulus of fibers is assumed to be constant,
and the disorder of the material is captured solely by assigning random breaking thresholds 
to the fibers. However, it is well known that composite materials made up
as a mixture of several ingredients, can also have a higher degree of disorder where not only
the strength but even the Young modulus of local material elements can vary 
\cite{pradhan_failure_2010}. 

The goal of the present Letter is to capture the heterogeneity of the stiffness of elements 
and investigate the emerging complex fracture process in the framework of fiber bundle 
models. We focus on aspects relevant 
from the viewpoint of statistical physics and show that the
system has several interesting novel features. 
 Assuming a constant strength for fibers, 
under global load sharing the strain
is everywhere the same in the bundle, hence, the fibers break in the decreasing order of their stiffness
values. It has the consequence that the tail of the disorder distribution plays a crucial role
in the breaking process. For two component systems analytical calculations revealed that the macroscopic strength
of the mixture is always lower than the strength of its components.
As the main outcome, our investigations showed that in a bundle of
power law distributed stiffnesses a disorder induced transition occurs from a perfectly 
brittle phase where the first fiber breaking triggers the immediate collapse of the bundle to
a quasi-brittle phase where precursors emerge before failure. Based on analytical calculations 
and computer simulations we construct the phase diagram of the 
system and determine the critical exponents. 

\section{FBM with Random Young modulus}
In the classical FBM a bundle of $N$ fibers is considered which all
have a linearly elastic behavior characterized by the same 
Young modulus $E$. The fibers can sustain a finite load, i.e.\ when the load 
on them exceeds a threshold value $\sigma_{th}$, the fibers break irreversibly.
To capture the heterogeneity of the local physical properties of materials 
it is assumed that the strength of fibers $\sigma_{th}$ is a random variable
with a probability density $p(\sigma_{th})$ and a cumulative distribution 
$P(\sigma_{th})=\int_{0}^{\sigma_{th}}p(x)dx$. After a fiber breaks, its load
has to be overtaken by the remaining intact ones which introduces interaction between 
the fibers. For the load redistribution two limiting cases are usually analyzed:
equal load sharing (ELS) means that all the intact fibers share the same
load irrespective of their distance from the failed one 
\cite{sornette_prl_78_2140,kloster_pre_1997,raischel_pre_2006}. ELS implies that there is
no stress concentration in the bundle. In the opposite limit of localized load sharing,
the load is redistributed over the close vicinity of broken fibers leading to
strong overloads around failed regions \cite{raischel_pre_2006}. 

Assuming equal load sharing the macroscopic constitutive equation $\sigma(\varepsilon)$ 
of the system can easily be obtained analytically. Loading the bundle parallel to the fibers' 
direction, the strain $\varepsilon$ is everywhere the same in the system. 
At a given $\varepsilon$ all the intact fibers keep the load $E\varepsilon$ and their 
fraction is given by $[1-P(E\varepsilon)]$, since the fibers with breaking thresholds
$\sigma_{th} < E\varepsilon$ have already failed. It follows that the constitutive equation
takes the form
\begin{eqnarray}
\sigma(\varepsilon) = E\varepsilon\left[1-P(E\varepsilon) \right].
\label{eq:classic_const}
\end{eqnarray}
It has been shown in the literature that for a broad class of disorder distributions
$p(\sigma_{th})$ the constitutive curve $\sigma(\varepsilon)$ has a quadratic maximum whose
position and value define the critical deformation $\varepsilon_c$ and critical stress
$\sigma_c$ of the system, respectively \cite{pradhan_failure_2010,sornette_prl_78_2140,kloster_pre_1997,kun_basquin_prl2008,
hansen_crossover_prl,kovacs_mix_gls}.

In the present paper we modify the classical FBM by introducing randomness for
the Young modulus of fibers described by the probability density function $p(E)$  over the
interval $E_{min}\leq E \leq E_{max}$. For simplicity we assume that the strength of fibers 
is constant $\sigma_{th}=const$ so that the random Young modulus is the only 
source of disorder. 

\subsection{Constitutive equation}
In order to derive the constitutive equation we assume that the system is
loaded through rigid bars which ensure that the deformation $\varepsilon$ is everywhere 
the same in the bundle. The random Young modulus of fibers has the consequence that
at a given value of $\varepsilon$ the fibers carry different loads 
$\sigma_i= E_i\varepsilon$, 
where $E_i$ denotes the Young modulus of fiber $i$ $(i = 1,\ldots ,N)$.
Slowly increasing the externally imposed deformation $\varepsilon$ those
fibers will first reach the constant breaking threshold $\sigma_{th}$, 
which have  the highest Young modulus, i.e.\ for which 
$\sigma_{th}=E_i\varepsilon$ hold. It follows from the above arguments
that in the bundle fibers break in the decreasing order of their Young modulus
and at a given strain $\varepsilon$ those fibers are broken whose Young
modulus exceeds $\sigma_{th}/\varepsilon$.  

The load kept by the fibers having Young modulus between $E$ and $E+dE$
reads as $E\varepsilon p(E)dE$ which has to be summed up from $E_{min}$ to
the upper limit of intact fibers $\sigma_{th}/\varepsilon$. Hence, the
the constitutive equation of a fiber bundle with random Young
modulus and a constant breaking threshold can be cast into the generic form
\begin{eqnarray}
\sigma(\varepsilon) = \varepsilon \int _{E_{min}}^{\sigma_{th}/\varepsilon} E p(E)
dE.          
\label{eq:constit2}
\end{eqnarray}
To understand the mechanical 
behavior of the bundle it is instructive to consider some limiting cases of the
loading process:
at small deformations $\varepsilon \rightarrow 0$, the upper limit of integration in 
Eq.\ (\ref{eq:constit2}) goes to infinity so that we have
\begin{equation}
\sigma(\varepsilon \rightarrow 0)\approx \varepsilon \int_{E_{min}}^\infty p(E) E dE =
\left<E\right> \varepsilon,
\end{equation}
where $\left<E\right>$ denotes the average value of the Young modulus of fibers. 
The result implies that the system displays linear behavior in the limit of small strains, 
where the macroscopic Young modulus of the bundle is equal to the average Young 
modulus of the fibers.
At high deformations $\varepsilon \rightarrow \infty$ the upper 
limit of integration goes to zero so that $\sigma(\varepsilon \rightarrow \infty) = 0$
indicating the breaking of all fibers. 
Between the two limits, the constitutive curve can have a maximum whose position
$\varepsilon_c$ is obtained from the equation 
\begin{equation}
\left. \frac{d\sigma}{d\varepsilon}\right|_{\varepsilon_c} = \int_{E_{min}}^{\sigma_{th}/\varepsilon_c} p(E) E dE - 
\left( \frac{\sigma_{th}}{\varepsilon_c} \right)^2  p \left( \sigma_{th}/ \varepsilon_c
\right) = 0.        \label {complexmax}
\end{equation}
defining the critical strain of the bundle. Under stress controlled loading
the bundle fails catastrophically when reaching the maximum stress 
$\sigma_c=\sigma(\varepsilon_c)$, which is the macroscopic strength 
of the bundle.

\section{Mixture of two components}
Recently, it has been shown that forming a composite of hard and soft components 
can increase the strength of the material. One
example of such two-component composites where strengthening occurs is a so-called 
double-network (DN) gel, where a network of brittle polyelectrolyte
gel and flexible polymer chains are mixed \cite{dnetwork,japan}. 
As an application of our modeling approach, first we consider a simple two-component
mixture composed of two subsets of fibers with different Young modulus $E_1$ and $E_2$
where $E_1 < E_2$.
The fractions of the two subsets are $p_1$ and $p_2$, where $p_1+p_2 = 1$ holds. 
The main parameters of the model are $r\equiv E_2/E_1$, and $k\equiv p_2/p_1$ 
with ranges $1 \leq r \leq \infty$ and $0 \leq k \leq \infty$, respectively. 

Starting from the generic expression Eq. (\ref{eq:constit2}) it follows that
before the first breaking event the bundle has a linearly elastic behavior
with the constitutive relation
\begin{equation}
\sigma = \varepsilon (E_1 p_1 + E_2 p_2) = \varepsilon E_1 p_1 (1+rk),
\end{equation}
where the average Young modulus has the expression $\left<E \right>=p_1E_1+p_2E_2$.
Increasing the external load, first fibers of the subset with the higher Young modulus 
$E_2$ break. This breaking occurs instantaneously, i.e.\ all the fibers of the subset
are removed at once when $\sigma_{th}=E_2 \varepsilon$ is reached.
For the stress at the first breaking $\sigma_1$ we can write 
$\sigma_1 = \varepsilon_1 \left<E\right>  = \sigma_{th}\left(p_1/r + p_2\right)$. 

\begin{figure}
 \begin{center}
\epsfig{bbllx=5,bblly=560,bburx=480,bbury=770,file=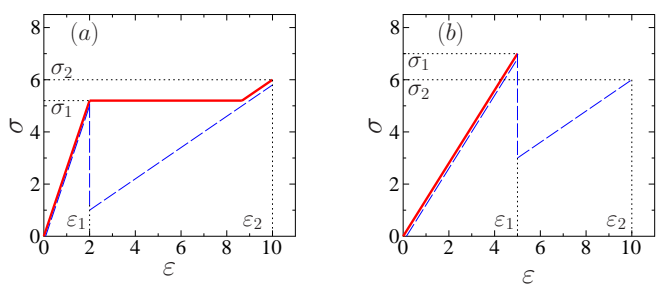,width=8.8cm}
 \caption{(Color online) Macroscopic response of the two-component system for 
two different values of $r$. The solid red line represents 
the constitutive curve under stress controlled loading while the blue dashed line 
shows the behavior under strain controlled loading.
The value of the parameters  $E_1=1, p_1=0.6, p_2=0.4, k=0.66,
\sigma_{th}=10$ are fixed. In $(a)$ for $r=5$ the stability condition Eq.\ (\ref{cond1}) 
holds so the system stabilizes after the first collapse and the macroscopic strength
of the bundle is determined by the second subset.
In $(b)$ for $r=2$ the stability condition fails so that macroscopic
failure occurs right at the breaking of the first subset.} 
\label{strainmacro}
  \end{center}
\end{figure}
Performing strain controlled loading, after the first collapse (of the fibers with the 
higher Young modulus $E_2$) the strain of the system $\varepsilon$ remains the same 
while the stress drops down as a result of the breaking by an amount 
$\Delta \sigma=\sigma_{th}p_2$ to the new value $\sigma'_1=\sigma_{th}p_1/r$.
After the stress drop, we can keep loading the system until 
we reach the second collapse, where all fibers with the lower
Young modulus $E_1$ break. This event occurs at the strain $\varepsilon_2=\sigma_{th}/E_1$ 
and stress $\sigma_2=\varepsilon_2\left<E'\right> = \sigma_{th} p_1$ values.
Comparing $\sigma_1$ and $\sigma_2$ we can check whether the maximum load capacity 
$\sigma_c$ of the system is determined by the collapse of the first or the 
second subset. Based on the above 
results it can be derived that if the parameters of the model $r$ and $k$ fulfill 
the condition
\begin{equation}
k<1-1/r,   \label{cond1}
\end{equation}
the critical load $\sigma_c$ of the system is determined by the
second collapse $\sigma_c=\sigma_2$, otherwise, by the collapse of the
first subset with the higher Young modulus $\sigma_c = \sigma_1$. 
\begin{figure}
 \begin{center}
\epsfig{bbllx=0,bblly=560,bburx=490,bbury=770,file=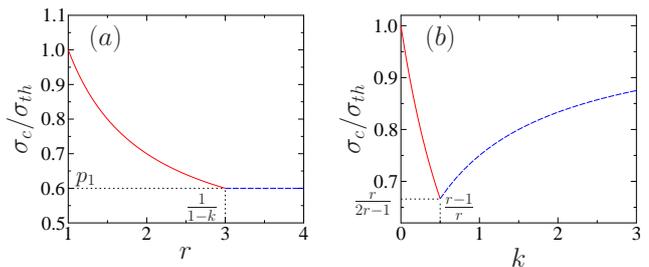,width=8.5cm}
 \caption{(Color online) Strength of the two-component system normalized by the breaking threshold
$\sigma_{th}$ as a function of $(a)$ the ratio $r\equiv E_2/E_1$ with fixed $k=2/3$ and
$(b)$ the ratio $k \equiv p_2/p_1$ with fixed $r = 2$. 
From both $(a)$ and $(b)$ it can be seen that the single component system is stronger
than the two component one.}        \label{sigvsr}
  \end{center}
\end{figure}

Under stress controlled loading conditions, the system follows the same constitutive curve
$\sigma(\varepsilon)$ as under strain controlled loading
until the first collapse. Assuming $\sigma_c=\sigma_2$, at this point the stress 
is kept constant and the bundle gets elongated
by an amount $\Delta \varepsilon=k\sigma_{th}/E_1$ to the new value 
$\varepsilon' =\sigma_{th}/E_2 + k\sigma_{th}/E_1$ 
where the remaining fibers will be able to
sustain the load.
Figure \ref{strainmacro} presents the macroscopic response of the two-component system
for two different parameter sets: in $(a)$ the condition $k<1-1/r$ holds so that
stability is retained after the failure of the stiffer subset of fibers, while in $(b)$
macroscopic failure occurs right after the first breaking. We indicated the path
of both the strain and stress controlled loading.

The most important characteristics of the system is the macroscopic strength $\sigma_c$.
Based on the above results we can express the ratio of the fracture strength 
of the bundle $\sigma_c$ and of the breaking threshold $\sigma_{th}$ as function
of the composition parameters $k$ and $r$.
For fixed values of $k$ we obtain
\begin{eqnarray}
 \sigma_c/\sigma_{th}(r) &=& \left\{  
\begin{array}{ll} 
\frac{p_1}{r}+p_2, & 1\leq r \leq \frac{1}{1-k}; \\
p_1, & \frac{1}{1-k} < r ,  \label{kfix}
\end{array}
\right.      
\end{eqnarray}
and for fixed $r$ the calculations yield
\begin{eqnarray}
 \sigma_c/\sigma_{th}(k) &=& \left\{  
\begin{array}{ll} 
1/(1+k), & 0 \leq k \leq 1- \frac{1}{r}; \\
\frac{1+rk}{r+rk}, & 1-\frac{1}{r} <k .
\end{array}  \label{rfix}
\right.            
\end{eqnarray}
The results are plotted in Fig.\ \ref{sigvsr}. The main outcome of the calculations is
that in the framework of the fiber bundle model with infinite range of interaction, i.e.\
global but not equal load sharing ensured by the rigid loading bars, 
the single component system has always higher strength $\sigma_c$ than a two-component
mixture. The worst situation with the lowest strength is obtained for the
parameter combination $k=1-1/r$. The results are in agreement with Ref.\ \cite{japan}
where it was shown that the strengthening in two component mixtures only occurs
if a large enough crack exists in the sample before the loading starts.  

\section{Disorder induced brittle to quasi-brittle transition}
As the next step of the investigations we consider the case when the Young modulus
of fibers has a continuous distribution $p(E)$ over the 
interval $E_{min} \leq E \leq E_{max}$. Starting from the generic form 
of the constitutive equation Eq.\ (\ref{eq:constit2}) 
it can be shown that for a uniform distribution of $E$ the breaking of the
first fiber with the highest Young modulus results in immediate catastrophic collapse
of the entire bundle. The reason is that for the uniform distribution a relatively large 
fraction of fibers have Young modulus in the vicinity of the largest value $E_{max}$, 
and hence, the load increment created by the first breaking can trigger a catastrophic
avalanche.  For the Weibull distribution such catastrophic collapse does not occur
due to the exponential tail of the distribution. These arguments show the importance
of the shape of the stiffness disorder in the vicinity of the upper bound, hence, 
in the following we consider a power law distribution $p(E) = C  E^{-a}$ for the Young modulus,
where the functional form can be controlled by varying the exponent $0 \leq a$. 
In the limiting case $a=0$ we recover the uniform distribution, 
while increasing $a$ the function $p(E)$ decreases faster making the majority of fibers
less stiff. 
\begin{figure}
 \begin{center}
\epsfig{bbllx=0,bblly=565,bburx=480,bbury=765,file=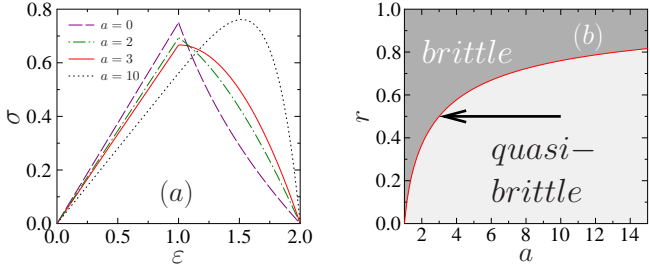,width=8.6cm}
\caption{(Color online) $(a)$ Macroscopic behavior of the bundle with power law distributed
Young modulus for different values of the exponent $a$ and fixed values of 
 $E_{min}=0.5$ and $E_{max}=1$. $(b)$ Phase diagram of the fiber bundle with power law distributed Young modulus
on the $r$-$a$ plane. The phase boundary between the perfectly brittle and quasi-brittle
macroscopic behavior is formed by the curve of $r_c(a)$.}   \label{expmacro}
  \end{center}
\end{figure}

The constitutive equation $\sigma(\varepsilon)$ of the system can be determined 
by substituting the distribution $p(E)$ after proper normalization into the generic form of 
Eq.\ (\ref{eq:constit2})
\begin{eqnarray}
\sigma_a(\varepsilon) &=& \left[ \frac{1-a}{2-a} \
\frac{1}{E_{max}^{1-a}-E_{min}^{1-a}} \right]
\left(\sigma_{th}^{2-a} \ \varepsilon^{a-1} - E_{min}^{2-a} \ \varepsilon \right)  
\label{expmacrogen}, \nonumber \\
\sigma_1(\varepsilon) &=&  \frac{1}{\ln {\frac{E_{max}}{E_{min}}}} \left( 
\sigma_{th} - E_{min} \ \varepsilon \right),   \\
\sigma_2(\varepsilon) &=& \frac{E_{max}E_{min}}{E_{max} - E_{min}} 
\left[ \ln\left({\frac{\sigma_{th}}{E_{min}}}\right) \ \varepsilon -
\ln{(\varepsilon}) \ \varepsilon \right],\nonumber
\end{eqnarray}
where $\sigma_a(\varepsilon)$,  $\sigma_1(\varepsilon)$, $\sigma_2(\varepsilon)$  correspond 
to a generic $a$ different from 1 and 2, to the special case of $a=1$ and $a=2$, respectively.
Figure \ref{expmacro}$(a)$ shows representative examples of the constitutive curve $\sigma(\varepsilon)$ 
for different values of the exponent $a$ with fixed $E_{min} = 0.5$, $E_{max} = 1$. 
It can be observed that at low values of $a$ the constitutive curve $\sigma(\varepsilon)$ 
has a sharp maximum, i.e.\ $\sigma(\varepsilon)$ has a linear behavior
up to the maximum followed by immediate collapse which indicates a perfectly brittle behavior. 
However, at high $a$ values $\sigma(\varepsilon)$ develops 
a quadratic maximum. i.e.\ nonlinear behavior occurs before the maximum is reached.
It follows that in this parameter range, where the stiffness distribution
decreases rapidly, the system becomes quasi-brittle being able
to suffer several avalanches of fiber breaks before collapsing in a catastrophic avalanche.
In this case the macroscopic failure of the system is preceded by precursors which
provide important signals of the imminent failure event.
The boundary of the two phases can be found by analyzing
the derivative of the constitutive curve $d\sigma/d\varepsilon$ at the point of
the first fiber breaking $\varepsilon=\sigma_{th}/E_{max}$. Differentiating Eq.\ 
(\ref{expder1}) we get
\begin{eqnarray}
\left. \frac{d\sigma_a}{d\varepsilon}\right|_{\frac{\sigma_{th}}{E_{max}}} &=&  
E_{max} \frac{1-a}{2-a} \left[ \frac{a-1-r^{2-a}}{1-r^{1-a}} \right],   \label{eq:der_gen}\\ 
\left. \frac{d\sigma_1}{d\varepsilon}\right|_{\frac{\sigma_{th}}{E_{max}}}  &=&  
E_{max}\frac{r}{\ln{r}}, \label{expder1} \\
\left. \frac{d\sigma_2}{d\varepsilon}\right|_{\frac{\sigma_{th}}{E_{max}}}  &=&
E_{max}\frac{r}{r-1} \left[ 1 + \ln{r} \right],\label{expder2}
\end{eqnarray}
where the parameter $r=E_{min}/E_{max}$ can vary in the range $0<r\leq 1$. If the derivative is negative 
the bundle is perfectly brittle and 
collapses without precursors, if however it is positive $d\sigma/d\varepsilon|_{\sigma_{th}/E_{max}} > 0$ 
precursors can be observed and
a quasi-brittle behavior emerges (see also Fig.\ \ref{expmacro}). Analytic calculations 
show that the stability features are determined by the range of Young modulus 
$r$ and by the exponent $a$:
\begin{figure}
 \begin{center}
\epsfig{bbllx=5,bblly=565,bburx=300,bbury=770,file=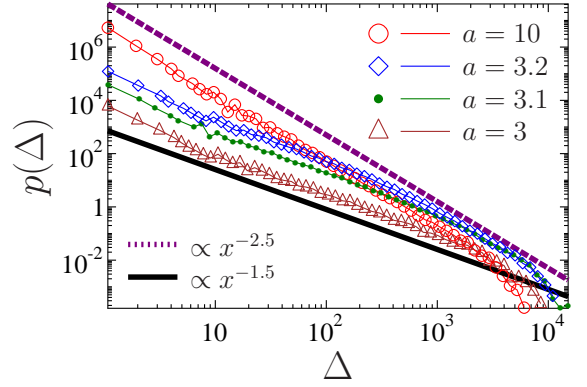,width=7.5cm}
\caption{(Color online) The burst size distribution for power law 
decreasing stiffness disorder for $10^6$ fibers averaged over $1000$ samples 
with different $a$ values and a fixed $r = 0.5$. Approaching the phase boundary
$a\to a_c=3$ a crossover occurs from the exponent 5/2 to a lower one 3/2.}
\label{burstexp}
  \end{center}
\end{figure}
For the interval $0\leq a \leq 1$ the system collapses at the instant 
of the first fiber breaking regardless of the value of $r$.
This also means that for uniformly distributed Young
modulus $a=0$ no stability can be obtained. 
For $a > 1$ the stable regime of $r$ is $r\geq r_c$ where the critical value $r_c$ 
as a function of $a$ can be cast in the form
\begin{equation}
r_c(a) = (a-1)^{1/(2-a)}.   \label{rcrit}
\end{equation} 
The result implies that at a given value of the exponent $a$ the range of Young modulus 
values has to be small enough to stabilize the system, or if we fix the range $r$ 
the distribution has to decay fast enough to obtain stability.
The results are summarized in Fig.\ \ref{expmacro}$(b)$ which provides a phase
diagram of the system on the $r$-$a$ plane. The boundary between the phases
of perfectly brittle and quasi-brittle behaviors is provided by the function $r_c(a)$
Eq.\ (\ref{rcrit}).

\subsection{Microscopic dynamics - statistics of avalanches} \label{stata} 
Under stress controlled loading, after each breaking 
event the load of the broken fiber gets redistributed over the intact ones which can induce
additional breakings and finally can even trigger an entire avalanche of failing fibers.
The size of the avalanche $\Delta$ is defined as the number of fibers breaking in the avalanche.
The emergence of avalanches is the direct consequence of the quasi-brittle macroscopic
response in FBMs which can also be used to forecast the imminent failure event. 
It is important to emphasize that in our FBM where the fibers have random Young modulus,
the load carried by the fibers is different even if we assume global load sharing. 

\begin{figure}
 \begin{center}
\epsfig{bbllx=5,bblly=565,bburx=490,bbury=770,file=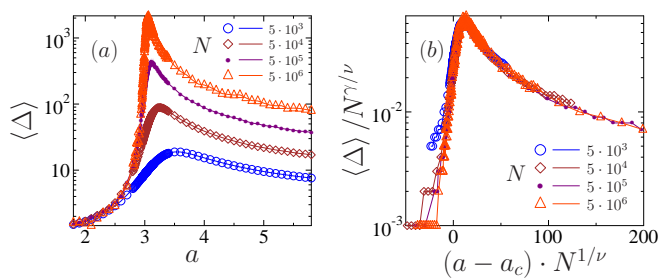,width=8.5cm}
\caption{(Color online) $(a)$ Average avalanche size as a function of $a$ for several different
system sizes $N$. As $N$ increases the position of the maximum $a_c(N)$ shifts towards
the critical point of the infinite system $a_c(\infty)$ determined analytically.  
$(b)$ Finite size scaling of the average cluster size. The good 
quality collapse was obtained with the exponents $\gamma=2$ and $\nu=3$.}     
\label{burstaver}
  \end{center}
\end{figure}
In order to study the microscopic avalanche dynamics computer simulations were carried 
out at a fixed range of Young modulus with $r=0.5$
which yields a critical value of $a_c=3$ for the infinite system from Eq.\ (\ref{rcrit}). 
To characterize the statistics of avalanches we determined their size distribution
$p(\Delta)$ for several values of the exponent $a$ inside the stable 
regime between the limits $a_c\leq a \leq 10$ as it is represented by the arrow 
in Fig.\ \ref{expmacro}$(b)$. It can be observed in Fig.\ \ref{burstexp} that the burst size 
distribution has a power law behavior 
\begin{eqnarray}
p(\Delta) \sim \Delta^{-\tau}
\end{eqnarray}
for all the parameter values considered. It has to be emphasized that as we approach
the phase boundary $a_c$ from the quasi-brittle 
phase, the distribution $p(\Delta)$ shows a crossover from a higher to a lower exponent:
Far from the phase boundary the value $\tau=5/2$ is obtained in agreement with the usual 
exponent of FBMs with equal load sharing 
\cite{pradhan_failure_2010,kloster_pre_1997,kovacs_mix_gls}. 
As the response becomes more brittle $a\to a_c(r)$ 
the exponent switches to a lower value $\tau=3/2$. This crossover is similar to what 
is observed in classical FBMs with solely strength disorder when increasing 
the lower bound of the range of strength values
towards the critical one of immediate collapse \cite{hansen_crossover_prl,raischel_pre_2006}. 
However, in our case the range of Young 
modulus $r$ is fixed, while the functional form of the disorder distribution is changed. 
The results indicate that the higher degree of brittleness implies a lower exponent $\tau$
of the burst distribution due to the higher frequency of avalanches of larger size.

In order to reveal the nature of the transition from the quasi-brittle phase to the 
one of perfectly brittle behavior with catastrophic collapse, we extended the simulations
to $a$ values inside the brittle phase. We determined the average size of avalanches 
$\left<\Delta\right>$ as a function of $a$ for several different system sizes $N$. 
The average size of bursts $\left<\Delta\right>$ is calculated as the 
ratio of the second and first moments of the burst size distribution skipping the largest
avalanche
\begin{eqnarray}
\left<\Delta\right> = \frac{\sum_i \Delta_i^2}{\sum_i \Delta_i}.
\end{eqnarray}
It can be observed in Fig.\ \ref{burstaver}$(a)$ that the average avalanche size has a 
maximum as $a$ is varied. Increasing the system size $N$, the maximum gets more and more
peaked and its position shifts to lower $a$ values. Left from the maximum $\left<\Delta\right>$ 
decreases rapidly and goes to zero marking the perfectly brittle response in this $a$ regime.
It follows that the position of the maximum of $\left<\Delta\right>$ can be identified as the
pseudo-critical point $a_c(N)$ of the finite size system which converges towards the 
critical point of the infinite system $a_c(\infty)$ determined analytically in the previous section. 
Assuming that the system has a continuous phase transition from the quasi-brittle to the brittle
phase when $a$ decreases, the finite size scaling form 
\beq{
a_c(N) = a_c(\infty) + BN^{-1/\nu}
\label{eq:hyper}
}
has to hold, where $\nu$ is the correlation length exponent of the
transition \cite{percol}. 
Figure \ref{burstaver}$(b)$ presents that by rescaling the curves of $\left<\Delta\right>(a,N)$ by
an appropriate power of the system size $N$ taking into account the functional form Eq.\ (\ref{eq:hyper})
of $a_c(N)$, the curves of Fig.\ \ref{burstaver}$(a)$ can
be collapsed on top of each other. The high quality collapse implies that 
$\left<\Delta\right>(a,N)$ has the scaling structure
\begin{eqnarray}
\left<\Delta\right>(a,N) = N^{\gamma/\nu}F^{(1)}((a-a_c)N^{1/\nu}),
\end{eqnarray}
where $\gamma$ is the susceptibility exponent of the brittle to quasi-brittle transition
and $F^{(1)}$ denotes the scaling function \cite{percol}. 
Based on the data collapse analysis the values $\gamma=2$ and $\nu=3$ were obtained.

Perfectly brittle behavior of the system implies that no damage can be accumulated 
prior to failure. However, in the quasi-brittle phase a finite $N_b$
number of fibers break in avalanches before the catastrophic avalanche. 
Hence, we introduce the order parameter of the system as
the fraction of fibers breaking prior to failure $p_b=N_b/N$ which is zero in the 
brittle and has a
finite value in the quasi-brittle phase. For the infinite system $p_b$ can be obtained analytically
as $p_b = 1-P(\sigma_{th}/\varepsilon_c)$, 
where $P$ is the cumulative distribution of the Young modulus and $\varepsilon_c$ is the
critical deformation of the system obtained from Eq.\ (\ref{complexmax}). 
Finally, $p_b$ can be cast into the form
\beq{
p_b = \frac{r^{a-1} - (a-1)^{(a-1)/(2-a)}}{r^{a-1} -1},
\label{eq:order}
}
\begin{figure}
 \begin{center}
\epsfig{bbllx=5,bblly=450,bburx=350,bbury=750,file=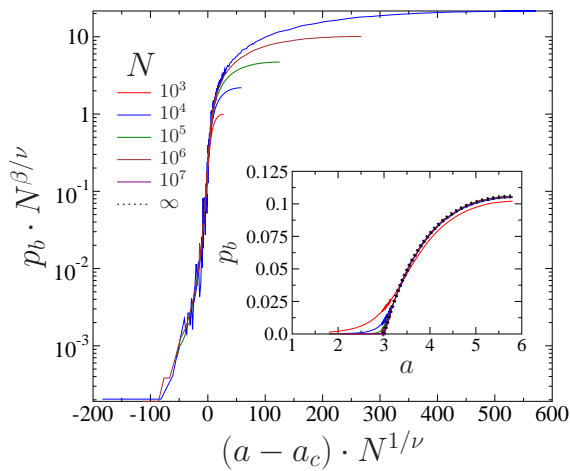,width=7.5cm}
\caption{(Color online) Inset: fraction of fibers $p_b$ breaking before final 
failure as a function of $a$ for different system sizes together with the analytic
solution Eq.\ (\ref{eq:order}). Main panel: Finite size scaling of $p_b$. The good 
quality collapse was obtained with $\beta=1$ and $\nu=3$.}     \label{fig:order}
  \end{center}
\end{figure}
which has a power law dependence on the distance from the critical point 
$p_b\sim (a-a_c(r))^{\beta}$.
The value of the order parameter exponent is $\beta=1$. The inset of Fig.\ \ref{fig:order}
presents $p_b$ as a function of $a$ obtained by simulations for different system sizes 
together with the analytic solution Eq.\ (\ref{eq:order}). The main panel of the figure
demonstrates the good quality collapse of $p_b(a,N)$ assuming the scaling structure
\beq{
p_b(a,N) = N^{-\beta/\nu}F^{(2)}((a-a_c)N^{1/\nu}).
}
The best collapse was obtained with the exponents $\beta=1$ and $\nu=3$ showing the 
consistency of the results. The scaling function is denoted by $F^{(2)}$. It is 
interesting to note that the critical exponents $\beta, \gamma$, and $\nu$ fulfill
the hyper-scaling relation $\beta+\gamma=\nu$.

\section{Summary}
We investigated the fracture process of heterogeneous materials in the framework of 
a fiber bundle model. As a novel feature of the model we assumed that the source of
disorder is the randomness of the Young modulus of fibers, while their breaking strength
is kept constant. 
We showed by analytical and numerical calculations that the system exhibits several
interesting features both on the micro and macro scales. 
For two component systems we demonstrated that the strength of the mixture is always 
lower than that of the individual components.

To obtain a deeper understanding of the breaking
process, a power law distribution was introduced for the Young modulus defined over a finite 
range. We showed analytically that varying the range of the Young modulus and 
the value of the exponent of the disorder distribution the system exhibits a transition from
a perfectly brittle phase where the breaking of the first fiber triggers the immediate
collapse of the system to a quasi-brittle one where macroscopic failure is preceded by avalanches
of breaking events and by a non-linear macroscopic response. Based on analytic 
calculations and computer simulations we
constructed the phase diagram of the system and we showed that the brittle to 
quasi-brittle transition occurs analogously to continuous
phase transitions. The critical exponents of the transition, determined 
by finite size scaling analysis, fulfill the hyper-scaling relation 
$\beta+\gamma=\nu$. It is a crucial feature of the system that due to the global load
sharing there is no spatial correlation between fiber breakings, 
i.e.\ broken fibers nucleate completely randomly all over the bundle. This is the reasons
why the brittle to quasi-brittle transition shows some similar features to percolation. 
However, it has to be emphasized that the macroscopic failure of the bundle has nothing
to do with the appearance of a spanning cluster of broken fibers which is also
indicated by the fact that the fraction of broken fibers $p_b$ at global failure is
relatively small.    
For future investigations, it would be very interesting
to consider the competition of two sources of disorder, i.e.\ to investigate the case
where both the Young modulus and the breaking strength of 
fibers are random. Work in this direction is in progress. 

\acknowledgments
The work is supported by T\'AMOP 4.2.1-08/1-2008-003 project. The
project is implemented through the New Hungary Development Plan,
co-financed by the European Social Fund and the European Regional
Development Fund.
F.\ Kun acknowledges the Bolyai Janos fellowship of HAS.

\end{document}